\documentstyle[12pt]{article}
\begin{document}

\title{Divergences in the Effective Action for Acausal Spacetimes}
\author{Michael J. Cassidy \\ Department of
  Applied Mathematics and Theoretical Physics \\ University of
Cambridge, Silver St., Cambridge CB3 9EW, England}
\date{\today \\DAMTP/R-97/24}
\maketitle

\baselineskip=20pt

\begin{abstract}
The 1--loop effective Lagrangian for a massive scalar field on an arbitrary causality
violating spacetime is calculated using the methods of Euclidean quantum field theory
in curved spacetime. Fields of spin ${1\over2}$, spin 1 and twisted
field configurations are
also considered. In general, we find that the Lagrangian diverges to minus infinity 
at each of the $n$th polarised
hypersurfaces of the spacetime with a structure governed by a DeWitt--Schwinger type
expansion.  
\end{abstract}

\pagebreak

\section{Introduction}
\label{intro}

If one attempts to quantise fields on acausal spacetimes, one inevitably runs into
awkward problems of interpretation. Quantities that are well defined in globally
hyperbolic spacetimes can become ambiguous in geometries where strong causality is
violated. For example, the first attempt to construct an interacting quantum field 
theory on a
Morris, Thorne, Yurtsever \cite{thorne} type wormhole spacetime found that
the $S$ matrix was nonunitary when the state evolved through the region of
closed timelike curves (CTCs) \cite{fps}.

One also encounters problems with the definition of a suitable Green function.
Consider the (normally well defined) commutator of two free field operators $iD(x,y) =
[\phi(x),\phi(y)]$. In an acausal spacetime, even if $x$ and $y$ are locally spacelike
separated, it is not clear that $D=0$ because there may be a large timelike loop
connecting the two points, due to the nontrivial homotopy of the spacetime. 

Various attempts to circumvent these problems in a consistent way have been suggested
\cite{2,3,4,5,6,7}. This paper is concerned with the Euclidean approach, proposed in a
recent paper by Hawking \cite{2}. Motivation for this proposal comes from the simple
observation that in Euclidean space, there are no CTCs and in particular, no
closed or self--intersecting null geodesics. Therefore, if one considers a Euclidean
space $M_E$ which has some acausal Lorentzian section $M_L$, then the appropriate 
analytic
continuation of quantities that are well defined on $M_E$ should give unambiguous
results valid on the chronology violating section.

 The object of this paper is to apply the methods of
Euclidean quantum field theory in curved spacetime to derive a 1--loop effective
action for fields of arbitrary mass and spin in a typical causality violating
spacetime. Accordingly, section\ \ref{geom} is devoted to a discussion of multiply 
connected
Euclidean spaces and their universal covering spaces. Section\ \ref{heat} reviews all
the necessary theory of the heat operator, most notably its divergence structure and
asymptotic expansion for multiply connected spacetimes. This expansion is then used in
section\ \ref{act} to derive the 1--loop effective Lagrangian for a massive scalar
field, renormalised by the point splitting method. The
corresponding expressions for twisted configurations and fields of spin ${1\over2}$ 
and spin 1 are also obtained. In section\ \ref{egs}, these results are applied to a 
number of interesting chronology violating spacetimes, including
Grant's generalisation of Misner space \cite{grant} and the wormhole spacetime studied 
by Kim and Thorne \cite{kth}. The relevance of these
results for chronology protection \cite{cpc} is discussed in section\ \ref{conc}.

\section{Multiply connected Euclidean spaces}
\label{geom}

Consider an arbitrary multiply connected Euclidean spacetime, $M_E$. This spacetime is
just the quotient space
\begin{equation}
M_E = {{\overline M}_E\over\Gamma}
\end{equation}
where ${\overline M}_E$ is the simply connected universal covering space and $\Gamma$
is a properly discontinuous, discrete group of isometries of ${\overline M}_E$.
$\Gamma$ is isomorphic to the fundamental group of $M_E$, $\pi_1(M_E)$ and $M_E$ is
obtained from ${\overline M}_E$ by identifying points equivalent under $\Gamma$. If
$\pi_1(M_E)=Z_\infty$, then the fundamental domains ({\it i.e.} the copies of $M_E$ in
${\overline M}_E$) can be labelled by a single integer $n$, usually
interpreted as a winding number. Copies of the point $p\in M_E$ in the covering space 
are labelled  ${\overline p}_n\in 
{\overline M}_E$, where the points ${\overline p}_n$ are obtained by repeated
application of $\Gamma$ to the right, {\it i.e.} ${\overline p}_n={\overline p}_0
\gamma_n$. 

It will be useful to have a concrete example to refer to throughout the paper.
Therefore, we shall consider the Euclidean section of Grant space
\cite{grant}. Grant space is just flat Minkowski space with points identified under a
combined boost in the $(x,t)$ plane and translation in the $y$ direction. It is
 the universal covering space of the Gott spacetime \cite{gott}, which describes 
two cosmic strings passing each other with a constant velocity. The appropriate
Euclidean section is flat Euclidean space with points identified under a combined
rotation plus a translation in the orthogonal direction, as before.
In other words, for an arbitrary point ${\overline q}=(\tau, r,\theta,z)$, the effect 
of acting on ${\overline q}$ by $\gamma_n$ is just
\begin{equation}
{\overline q}_n={\overline q}\gamma_n=(\tau+n\beta, r, \theta + n\alpha,z)\space.
\end{equation}
One recovers the Lorentzian Grant space by analytically continuing the rotation
parameter to a boost ($\alpha\rightarrow a=i\alpha$).

A fundamental quantity of interest is
$\sigma_n(p,\lbrace\beta_E\rbrace)=\sigma({\overline p},
{\overline p}_n)$, which gives the squared distance along the spacelike geodesic
connecting $p\in M_E$ to itself with winding number $n$. $\lbrace\beta_E\rbrace$ 
collectively denotes 
various metric parameters, which relate equivalent points in the covering space.
On the Euclidean section of Grant space, one would have
\begin{equation}
\sigma({\overline q},{\overline q}_n{}')= (\tau-\tau'-n\beta)^2 + r^2 + r'^2 -
2rr'\cos(\theta -\theta'-n\alpha) + (z-z')^2
\end{equation}
so that
\begin{equation}
\label{mis}
\sigma_n(q,\lbrace\alpha,\beta\rbrace)= 2r^2\biggl(1-\cos(n\alpha)\biggr) + n^2
\beta^2\space.
\end{equation}

On the Euclidean section $M_E$, provided the parameters $\lbrace\beta_E\rbrace$ are
nonzero, the equation 
\begin{equation}
\label{hyp}
\sigma_n(p,\lbrace\beta_E\rbrace)=0
\end{equation}
 can (in general) only be
satisfied if $n=0$. However, if one considers analytically continuing any of the
metric parameters to imaginary values, thus obtaining a Lorentzian spacetime $M_L$
with parameters $\lbrace\beta_L\rbrace$,
then one may be able to find solutions to the equivalent of (\ref{hyp}) for all values 
of $n$. In that
case, the point $p$ would be joined to itself by a (self--intersecting) null geodesic
with winding number $n$. One can define the $n$th polarised hypersurface as the set of
points $\lbrace p\in M_L: \sigma_n(p,\lbrace\beta_L\rbrace)=0\rbrace$. The Cauchy
horizon is just the limit of this family of surfaces as $n\rightarrow\infty$. One may
easily verify that by setting equation (\ref{mis}) equal to zero and then analytically 
continuing
$\alpha\rightarrow a=i\alpha$, one obtains the criterion for polarised hypersurfaces in
Grant space (see \cite{grant}).

\section{The Heat Operator}
\label{heat}

As a first step towards obtaining the effective Lagrangian, we need to examine the
structure of the heat operator defined on $M_E$. Heat operators on Riemannian
manifolds have been extensively studied, so here we only review the most relevant
properties. For further technical details, the reader is referred to the paper by Wald
\cite{wald} and its associated references. Quantum field theory on multiply connected
spacetimes is discussed in Dowker \cite{dowk1} and Dowker and Banach \cite{dowk2}.

We begin by considering the `wave operator' ${\cal{A}} =-\nabla^2 + m^2$, defined on the
dense domain $C_0^\infty(M_E)$ of smooth ($C^\infty$) functions of compact support.
$\cal{A}$ is a symmetric operator on $L^2(M_E)$, the Hilbert space of square
integrable functions on $M_E$. However, to do quantum theory we need a self--adjoint
operator so we must extend this domain of definition in an appropriate way. Since
$\cal{A}$ is positive on its initial domain, standard theory states that positive
self--adjoint extensions must exist. The only problem is that $\cal{A}$ is unbounded,
so there may be more than one possible extension. However, if $M_E$ is a complete
manifold without boundary, then Gaffney \cite{gaff} has shown that $\cal{A}$ has a
unique self--adjoint extension, defined as the closure of $\cal{A}$ and denoted by
$A$. The domain of $A$ is just the Cauchy completion of dom($\cal{A}$) in the norm
$\parallel\psi\parallel^2 + \parallel {\cal A}\psi\parallel^2$, for $\psi\in L^2(M_E)$.
 This
property of the wave operator is known as essential self--adjointness and also holds
for some incomplete manifolds, such as Euclidean space with a point removed. It does
not hold for manifolds with boundaries or most manifolds with singularities. In this
paper, we shall always assume that $\cal{A}$ is essentially self--adjoint.

The heat operator is defined as
\begin{equation}
e^{-\tau A} = \int e^{-\tau\lambda} dE_\lambda
\end{equation}
where $E_\lambda$ is the spectral family of $A$. Once $e^{-\tau A}$ has been
constructed, one can apply the functional calculus of self--adjoint operators
\cite{reed} to obtain mathematically well--defined expressions for quantities of
physical interest. If one considers the 1--parameter family of integrals
\begin{equation}
\label{gen}
K(s)=\int^\infty_0e^{-\tau A} \tau^{s-1} d\tau,
\end{equation}
then $K(1)$ and $K(0)$ are particularly interesting. $K(1)=A^{-1}$ defines the Feynman
propagator and $K(0)$ is related to $\ln A$ (the effective Lagrangian) by
\begin{equation}
\label{efflag}
\ln A=\lim_{\epsilon\rightarrow 0}\left( -\int^\infty_\epsilon e^{-\tau A}
{d\tau\over\tau} + (\gamma -\ln\epsilon)I\right)
\end{equation}
where $I$ is the identity operator and Euler's constant $\gamma = \int_0^\infty
e^{-x}\ln x dx$.

It is well known that for $\tau>0$, the heat operator is given by a smooth integral
kernel $H(\tau,x,x')$. Thus the only possible divergences in (\ref{gen}) which could
prevent $K(s)$ from being given by a smooth integral kernel $K(s,x,x')$ are those
which could arise as $\tau\rightarrow\infty$ (infra--red divergences) or
$\tau\rightarrow 0$ (ultra--violet). Infra--red divergences can only occur if the
field mass $m=0$ and $M_E$ is noncompact. Since we are considering the massive scalar
field, we shall not worry about these divergences. We shall be more concerned with the 
ultra--violet divergence
structure, which is completely determined by the asymptotic expansion of
$H(\tau,x,x')$ about $\tau=0$
\begin{equation}
\label{expan}
H(\tau,x,x')= {\Delta^{1\over2}(x,x')\over (4\pi\tau)^{d\over2}} e^{-m^2\tau}
e^{-{\sigma(x,x')\over4\tau}} \sum_{j=0}^N a_j(x,x') \tau^j
\end{equation}
The coefficients $a_j(x,x')$ are recursively obtained and depend on local geometric
quantities. $\sigma(x,x')$ was defined earlier as the square of the geodesic distance 
between $x$ and $x'$ and $d={\rm dim}(M)$. $\Delta(x,x')=-{\rm
det}(-\sigma_{;\mu\nu'})$ is the Van--Vleck determinant.

One can see that if $x\ne x'$, the factor $e^{-{\sigma(x,x')\over4\tau}}$ ensures that
$H$ vanishes as $\tau\rightarrow 0$ faster than any power of $\tau$. Therefore,
provided there are no infra--red divergences, $K(s)$ is given by an integral kernel
$K(s,x,x')$ which can only be singular when $x=x'$. 

In a multiply connected spacetime, one can express $H(\tau,x,x')$ in terms of
${\overline H}$, the heat kernel defined on the universal covering space. The most
general relation is
\begin{equation}
H(\tau,x,x')=\sum_\gamma a(\gamma){\overline H}(\tau, x, x'\gamma)
\end{equation}
where $a(\gamma)$ is a unitary, 1--dimensional representation of $\Gamma$ ({\it i.e.}
$a(\gamma_1)a(\gamma_2)=a(\gamma_2 \gamma_1)$). Note that from now on, points in the
covering space have no bars on them as the distinction between $x\in M_E$ and $x\in
{\overline M}_E$ should be clear. If $\Gamma=\pi_1(M_E)=Z_\infty$, one
can write
\begin{equation}
H(\tau,x,x')=\sum_{n=-\infty}^\infty a(\gamma_n) {\overline H}(\tau, x,
 x_0'\gamma_n)
\end{equation}
where $a(\gamma_n)=e^{2\pi in\delta}$ and $0\le\delta\le{1\over2}$. In general, for
real fields $a(\gamma_n)$ must be real also, so that $a(\gamma_n)$ can only take the
values $\pm 1$, where the negative value would correspond to a twisted field
configuration \cite{isham,dewish}. For the moment, we take $a(\gamma_n)=1$ for 
simplicity, and note that this decomposition can be suitably extended to the family of
integrals $K(s)$, so that
\begin{equation}
K(s,x,x')=\sum_{n=-\infty}^\infty {\overline K}(s, x,x_0'\gamma_n)
\end{equation} 

\section{The Effective Action} 
\label{act}

The effective action of the quantum field, $S$, is related to the operator $A$ by
\begin{equation}
e^{-S}=({\rm det}A)^{-{1\over2}}=e^{-{1\over2}{\rm tr}(\ln A)}.
\end{equation}
One would like to represent $\ln A$ by an integral kernel $L(x,x')$, so that
\begin{equation}
S={1\over2}\int L(x,x) g^{1\over2} d^4x\space.
\end{equation}
One could then obtain the energy--momentum tensor by functionally differentiating the
effective Lagrangian ${\cal L}(x)={1\over2}L(x,x)$ with respect to the metric
$g_{\mu\nu}$. However, from the above discussion of ultra--violet divergences in
$K(s,x,x')$, it is clear that $\ln A$
 will be singular at $x=x'$. We must therefore adopt some renormalisation prescription.

In the point--splitting approach \cite{chris1,chris2}, one first
considers the quantity
\begin{equation}
L(x,x')=-\int_0^\infty H(\tau,x,x') {d\tau\over\tau}
\end{equation}
which is well defined for $x\ne x'$. In 4 dimensions, the divergences which occur as
the limit $x'\rightarrow x$ is taken are governed entirely by the first 3 terms of the
asymptotic expansion (\ref{expan}). We therefore obtain a finite, renormalised
$L(x,x')$ by subtracting this divergent part from $L(x,x')$ before taking the
coincidence limit.

For a multiply connected spacetime $M_E$, we have
\begin{equation}
L(x,x')=-\sum_{n=-\infty}^\infty\int_0^\infty {\overline
H}(\tau,x,x_n'){d\tau\over\tau}.
\end{equation}
The $\tau$ integration is performed with the help of the definite integral \cite{grad}
\begin{equation}
\int_0^\infty x^{\nu -1} e^{-{\beta\over x} -\gamma x}dx =
2\left(\beta\over\gamma\right)^{\nu\over2} K_\nu\left(2\sqrt{\beta\gamma}\right)
\end{equation}
where $K_\nu$ is the modified Hankel function \cite{grad}. The contribution to
$L(x,x')$ from the first 3 terms in the series is
\begin{eqnarray}
L(x,x')=&-&{1\over(4\pi)^2} \sum_{n=-\infty}^\infty
\Delta^{1\over2}\left(x,x_n'\right)\Biggl[ 8
\left(m^2\over\sigma(x, x_n')\right)K_2\left(m\sqrt{\sigma
( x, x_n')}\right)\Biggr.\nonumber \\  &+& 
\Biggl. 4a_{1}(x,x'_n)\left(m^2\over\sigma_n\right)^{1\over2}
K_1\left(m\sqrt{\sigma_n}\right) 
+ a_{2n}\int_0^\mu e^{-m^2\tau-{\sigma_n\over4\tau}}{d\tau\over\tau} + O(\sigma_n)
\Biggr]
\end{eqnarray}
where the cutoff $\mu$ is to prevent infra--red divergences in the massless limit. The
first coefficient $a_0=1$ for scalar fields.

Recalling the discussion in section\ \ref{geom}, the only value of $n$ for which
$\sigma( x,x_n)\rightarrow0$ as $x'\rightarrow x$ is $n=0$. To
renormalise therefore, one drops the $n=0$ term in $L(x,x')$ and takes the coincidence
limit to obtain
\begin{eqnarray}
\label{final}
-2{\cal L}(x)&=&-L(x,x)= {1\over(4\pi)^2} \sum_{\scriptstyle n=-\infty\atop\scriptstyle 
n\ne0}^\infty \Delta^{1\over2}(x,x_n)\Biggl[
 8\left(m^2\over\sigma_n\right)K_2\left(m\sqrt{\sigma_n}\right)\Biggr.
\nonumber \\
&+&\Biggl. 4a_{1n}\left(m^2\over\sigma_n\right)^{1\over2} 
K_1\left(m\sqrt{\sigma_n}\right)
+a_{2n}\int_0^\mu e^{-m^2\tau-{\sigma_n\over4\tau}}{d\tau\over\tau} + O(\sigma_n).
\Biggr]
\end{eqnarray} 

It is now clear what happens if one can analytically continue any of the metric
parameters to obtain an acausal Lorentzian section. The quantity $\sigma(
x, x_n)$ goes to zero at each of the $n$th polarised hypersurfaces and 
hence the renormalised effective Lagrangian diverges. 

The first point to note is that equation (\ref{final}) should be understood as a
formal expression only. For the purposes of practical calculation, the first term
gives the Gaussian approximation \cite{bek} to the effective action which is only
 exact in special cases, namely flat space and the Einstein universe. In a more 
general spacetime, one has to consider the 
coefficients $a_i$ for $i>0$. The covariant expansion of these coefficients in terms
of the geodetic interval can be
found in the papers by Christensen \cite{chris1,chris2}. Consider the expansion of the
first nontrivial coefficient $a_1$ for a scalar field
\begin{equation}
a_1(x,x')= \left({1\over6}-\xi\right)R -
{1\over2}\left({1\over6}-\xi\right)R_{;\alpha}\sigma^{\alpha} +
\biggl(\dots\biggr)_{\alpha\beta}\sigma^{\alpha}\sigma^{\beta} +\dots
\end{equation}
where $\sigma_\alpha = \partial_\alpha \sigma$. 
Normally, one would have $\sigma^{\alpha}\rightarrow 0$ as the coincidence limit is
taken, which leaves a simple finite expression for the coefficient of interest. 
However, if identifications have been
made, the quantity $\sigma^{\alpha}$ does not necessarily go to zero as the
points are brought together. Hence we are left with an infinite number of terms which
may or may not converge. However, for most purposes one would only be interested in
the strongest divergence, which is given by the Gaussian approximation
\begin{equation}
-{\cal L}(x)={a_0\over2\pi^2}\sum_{n=1}^\infty \left(m^2\over\sigma_n\right)
\Delta_n{}^{1\over2}K_2\Bigl(m\sqrt{\sigma_n}\Bigr)\space.
\end{equation}

A twisted real scalar field configuration can be considered by including a factor of
$(-1)^n$ in the Lagrangian. In this case, the contributions from twisted and untwisted
fields cancel at odd numbered polarised hypersurfaces, but reinforce at even numbered
ones. 

For higher spin fields, the only factor which changes in this expression is the
coefficient $a_0$. For spin ${1\over2}$, $a_0$ is given by the unit spinor, whose
trace is just the number of spinor components ({\it i.e.} the dimension of the gamma
matrices used). For spin 1 fields, $a_0$ has four components and is just the metric
tensor $g_{\mu\nu}$ (in the Feynman gauge). The ghost Lagrangian is given by minus
twice the scalar Lagrangian, because one would have to consider two anticommuting
scalar ghost fields. Here, the ghost contribution would cancel with two of the vector
field components so overall, the spin ${1\over2}$ and spin 1 Lagrangians would still 
diverge to minus infinity at the polarised hypersurfaces.  

\section{Examples}
\label{egs}

In flat space, we can obtain an exact result. The Van--Vleck determinant 
$\Delta(x,x')=1$ and the only nonzero
coefficient is $a_0=1$, so for Euclidean space
identified under a combined rotation and orthogonal translation, one obtains
\begin{equation}
\label{masgr}
-{\cal L}(x)={1\over2\pi^2}\sum_{n=1}^\infty
\left(m^2\over\sigma_n\right)K_2\Bigl(m\sqrt{\sigma_n}\Bigr)
\end{equation}
which in the massless limit becomes
\begin{equation}
\label{malgr}
-{\cal L}(x)= {1\over\pi^2}\sum_{n=1}^\infty
{1\over\left(2r^2\Bigl(1-\cos(n\alpha)\Bigr) + n^2\beta^2\right)^2}.
\end{equation}
Analytically continuing $\alpha\rightarrow a=i\alpha$ yields the Grant space result
which as stated above, diverges at each of its polarised hypersurfaces.

The Gaussian approximation is also known to be exact in the Einstein universe, which
has topology $R\times S^3$. If one identifies points on the Euclidean section under a
combined rotation plus translation, then the effective Lagrangian can be calculated as
before. In this case, however, one must also sum over contributions from geodesics
which loop around the three--sphere more than once, so one has to sum over two winding
numbers $n$ and $m$. If the metric is written as 
\begin{equation}
ds^2= d\tau^2 + r^2\Bigl( d\chi^2 + \sin^2\chi\left(d\theta^2 + \sin^2\theta
d\phi^2\right)\Bigr)
\end{equation}
and the points $(\tau,\chi,\theta,\phi)$ and $(\tau+m\beta,\chi,\theta,\phi+m\alpha)$
are identified, then the geodetic interval is given by 
\begin{equation}
\sigma_{nm}(x,x')= (\tau-\tau'-m\beta)^2 + (s_m + 2\pi nr)^2
\end{equation}
where
\begin{equation}
\cos\left(s_m\over r\right) = \cos\chi\cos\chi'+ \sin\chi\sin\chi'
(\cos\theta\cos\theta' + \sin\theta\sin\theta'\cos(\phi-\phi'-m\alpha))\space.
\end{equation}
One therefore obtains 
\begin{equation}
-L(x,x')= {1\over2\pi^2} \sum_{\scriptstyle m=-\infty\atop\scriptstyle 
m\ne0}^\infty \sum_{n=-\infty}^\infty {{s_m\over r}+2\pi n \over\sin\left(s_m\over
r\right)}{1\over\Bigl((\tau-\tau'-m\beta)^2 + (s_m + 2\pi nr)^2\Bigr)^2}
\end{equation}
for the integral kernel in the massless limit, where the factor $\left({s_m\over r} +
2\pi n\right)/\sin\left(s_m\over r\right)$ is the Van--Vleck determinant for this
spacetime. This expression can be written in an alternative form by combining terms of
positive and negative $n$. 
\begin{equation}
{1\over2\pi^2 r} \sum_{\scriptstyle m=-\infty\atop\scriptstyle 
m\ne0}^\infty \sum_{n=-\infty}^\infty {s_m\over\sin\left(s_m\over
r\right)}{y^4 + 2y^2(x+n)(x-n) + (x^2 + 3n^2)(x+n)(x-n)
\over16\pi^4 r^4(n+z_1)^2(n-z_1)^2(n+z_1^*)^2(n-z_1^*)^2}
\end{equation}
where the complex quantity
\begin{equation}
z_1=x+iy= {s_m + i(\tau-\tau'-m\beta)\over2\pi r}
\end{equation}
The sum over $n$ can be evaluated using the method of residues to obtain finally
\begin{equation}
-L(x,x')={1\over4\pi^2 r^4}\sum_{\scriptstyle m=-\infty\atop\scriptstyle 
m\ne0}^\infty {\left(\tau-\tau'-m\beta\over
r\right)^{-1}\sinh\left(\tau-\tau'-m\beta\over
r\right)\over\left(\cosh\left(\tau-\tau'-m\beta\over r\right)- \cos\left(s_m\over
r\right)\right)^2}
\end{equation}
 If one analytically
continues the parameter $\alpha\rightarrow a=i\alpha$ in this case, the spacetime that
 one obtains is the
product of three dimensional de Sitter space and the real line, periodically
identified under a combined boost and translation. The condition for polarised
hypersurfaces in this spacetime is given by
\begin{equation}
\cosh\left(m\beta\over r\right) -1 + \sin^2\chi\sin^2\theta\Bigl(1-\cosh(ma)\Bigr)=0
\space.
\end{equation}
This criterion and the Lagrangian both reduce to the Grant space expressions in the 
limit as $r\rightarrow\infty$, if one
defines a new radial coordinate by $r'=r\sin\chi\sin\theta$. 

One can also try to calculate the 
effective Lagrangian for the Anti--de Sitter analogue of Grant space. The Gaussian
approximation is exact for conformally invariant fields in this case also, due to the
fact that Anti--de Sitter space can be conformally mapped into half of the Einstein
static universe. The Euclidean
section of Anti--de Sitter space can be realised as the 4--dimensional hyperboloid 
\begin{equation}
-\left(\omega^0\right)^2 +\left(\omega^1\right)^2 + \left(\omega^2\right)^2 +
\left(\omega^3\right)^2 + \left(\omega^4\right)^2=r^2
\end{equation}
in the 5--dimensional space with metric
\begin{equation}
ds^2=-\left(d\omega^0\right)^2+\left(d\omega^1\right)^2 + \left(d\omega^2\right)^2 +
\left(d\omega^3\right)^2 + \left(d\omega^4\right)^2.
\end{equation}
If one defines
\begin{eqnarray}
\omega^0&=&{1\over r}\cosh\tau\sec\rho \nonumber \\
\omega^1&=&{1\over r}\tan\rho\cos\theta \nonumber \\
\omega^2&=&{1\over r}\tan\rho\sin\theta\cos\phi \nonumber \\
\omega^3&=&{1\over r}\tan\rho\sin\theta\sin\phi \nonumber \\
\omega^4&=&{1\over r}\sinh\tau\sec\rho
\end{eqnarray}
then the metric takes the form
\begin{equation}
ds^2={\sec^2\rho\over r^2}\left(d\tau^2 + d\rho^2 + \sin^2\rho\Bigl(d\theta^2 +
\sin^2\theta d\phi^2\Bigr)\right).
\end{equation}
Once again, we identify the points 
$(\tau,\rho,\theta,\phi)$ and $(\tau+n\beta,\rho,\theta,\phi+n\alpha)$. In Anti--de
Sitter space, the chief problem encountered when trying to construct quantum field
theoretic quantities comes from the fact that information can be lost to, or gained
from, spatial infinity in a finite coordinate time. Appropriate boundary conditions
need to be imposed at infinity, so that the field (or its gradient) vanishes there
\cite{ads}. If
one thinks of the Einstein universe as a cylinder, then Anti--de Sitter spatial
infinity is the timelike surface at $\chi={\pi\over2}$ obtained by slicing the
cylinder with a vertical plane wave. Thus, the Anti--de Sitter Lagrangian is obtained
from the Einstein expression by adding in the image charge at the antipodal point and
inserting the appropriate conformal weighting factor, to obtain
\begin{eqnarray}
-L(x,x')&=&{\cos^2\rho\cos^2\rho'\over4\pi^2}\sum_{\scriptstyle m=-
\infty\atop\scriptstyle m\ne0}^\infty\left[ {\left(\tau-\tau'-m\beta\over r\right)^{-1}
\sinh\left(\tau-\tau'-m\beta\over r\right)\over\left(\cosh\left(\tau-\tau'-m\beta\over
r\right)-\cos\left(s_m\over r\right)\right)^2}\right. \nonumber \\
&\pm&\Biggl. (\pi-\rho', \pi-\theta',
\pi+\phi')\hbox{ image charge}\Biggr]\space,
\end{eqnarray}
where the upper (lower) sign refers to Dirichlet (Neumann) boundary conditions.

As a final example, we consider a massless scalar field in the wormhole spacetime
originally studied by Kim and Thorne \cite{kth}, who calculated the (divergent)
behaviour of its renormalised energy--momentum tensor. One constructs this spacetime
by removing two 3--spheres of radius $b$ from Minkowski space and identifying the
resulting world tubes which form when one sets the right hand mouth moving towards the
left with speed $\beta$. Initially the two mouths are separated by a shortest 
distance $D$. Kim
and Thorne have calculated the Van--Vleck determinant and geodetic interval for this
spacetime. Combining their results with our expression, one immediately obtains
\begin{equation}
-{\cal L}(x)={1\over\pi^2}\sum_{n=1}^\infty {1\over D}\left(b\over2D\right)^{n-1}
{\xi^{2n}\left(1-\xi\right)\over
1-\xi^n}\left({1\over\lambda(x,x')}+{1\over\lambda(x',x)}\right)^2\space,
\end{equation}
for the Lagrangian, where $\xi=\left({1-\beta\over 1+\beta}\right)^{1\over2}$ is the 
inverse Doppler blueshift
suffered by a ray passing along the $X$ axis and through the wormhole, and the 
quantity $\lambda$ is defined by
\begin{equation}
\lambda(x,x')=2\left(b-\sqrt{b^2-\rho^2}\right) +X-T-(X'-T')\xi^n
\end{equation}
where a point $x$ has coordinates $(T,X,Y,Z)$ and $\rho=\sqrt{Y^2+Z^2}$ measures the
transverse shift of $x$ from the axis of symmetry.

\section{Discussion}
\label{conc}

One cannot dispute the fact that in many causality violating spacetimes, the
renormalised expectation value $\langle T_{\mu\nu}\rangle$ diverges as the Cauchy
horizon is approached. Indeed, the original chronology protection conjecture
\cite{cpc} was motivated heavily by this fact, and it was therefore proposed that the
back reaction induced by this divergent energy--momentum would distort the spacetime
geometry sufficiently to prevent the formation of CTCs. Recently, however, examples
have been presented which lead one to question the universal validity of this basic
mechanism and it is now known that $\langle T_{\mu\nu}\rangle$ does not necessarily
diverge for all initial quantum states as the Cauchy horizon is approached. Sushkov,
who considered automorphic fields on four dimensional Misner space \cite{sush}, gave
an example of a Hadamard state for which $\langle T_{\mu\nu}\rangle$ vanishes
everywhere on the initially globally hyperbolic region (see also Krasnikov
\cite{kras}). Actually, one does not even need to consider automorphic fields, as one
can readily find a simple counterexample from inspecting the closed form of the scalar
field energy--momentum tensor on Misner space, obtained by Euclidean methods. 
Recall that Misner space is just Minkowski space with points identified under a boost
in the $x$ direction. The appropriate Euclidean section of this spacetime, therefore,
is flat Euclidean space identified under a rotation, $\alpha$. This space also happens 
to be the analytic continuation of the Lorentzian
spacetime produced by an infinitely long cosmic string. The energy--momentum tensor
for a massless conformally coupled scalar field in the cosmic string spacetime is
well known, and is given on the Euclidean section (in ($\tau,r,\theta,z$) coordinates)
by
\begin{equation}
\langle T^\mu{}_\nu\rangle={1\over1440\pi^2 r^4}\left(\left(2\pi\over\alpha\right)^4
-1\right) {\rm diag}\Bigl(1,1,-3,1\Bigr)\space.
\end{equation}
Clearly, if one analytically continues the parameter $\alpha\rightarrow a=i\alpha$ in
this case, then the energy--momentum tensor vanishes everywhere if $a=2\pi$, so there
will be no divergence in this case. $\langle T_{\mu\nu}\rangle$ has also been shown to
be bounded at the Cauchy horizon for (sufficiently) massive fields in Gott space
\cite{boul} and Grant space \cite{tanhis}. Cramer and Kay \cite{ckay} have replied to
all of these examples by demonstrating that even though there is no divergence, 
$\langle T_{\mu\nu}\rangle$ must always be ill defined on the Cauchy horizon itself.
However, one is still left with the feeling that $\langle T_{\mu\nu}\rangle$ does not
quite tell the whole story.

In this paper, we have offered a new viewpoint by focusing on the effective Lagrangian
and a general expression for the leading order divergence at the polarised
hypersurfaces of a typical causality violating spacetime has been obtained.
Immediately one can apply this result to the examples outlined in the preceding
paragraph. In four dimensional Misner space, a quick inspection of (\ref{malgr}) with
the parameter $\beta=0$ shows that even though $\langle T_{\mu\nu}\rangle$ can remain
finite, ${\cal L}$ always diverges to minus infinity at the Cauchy horizon $r=0$.
Similarly, (\ref{masgr}) implies that ${\cal L}$ diverges at the Cauchy horizon in
Grant space (and therefore Gott space), even though $\langle T_{\mu\nu}\rangle$ can
remain finite for massive fields at the Cauchy horizon.

Finally, consider the behaviour of a Euclidean path integral of the form
\begin{equation}
\Psi=\int  {\cal D}[g]{\cal D}[\phi] e^{-S[g,\phi]}\space,
\end{equation}
where $S$ is obtained from a Lagrangian appropriate for some causality violating
spacetime. If the metric parameters are adjusted so as to introduce CTCs into the
spacetime, then we have already shown that the action diverges to minus infinity. If
one now interprets this path integral according to the no--boundary proposal, then it
seems that causality violations will be strongly enhanced, rather then suppressed. 
However, as one might expect, there is a subtlety
involved. We shall leave a full discussion of this problem to a future paper, but
conclude with a few brief remarks. Basically, one is interested in constructing the
density of states, or microcanonical partition function, as the squared amplitude
$\Psi^2$. The problem is that the microcanonical partition function should be defined
as a function of definite conserved quantities, such as energy and angular momentum.
The amplitude $\Psi$ above, however, is generally given as a function of the metric 
parameters
which relate equivalent points in the universal covering space, which could be inverse
temperature or angular velocity, for example. In order to achieve the correct result,
one must project the amplitude $\Psi$ on to states of definite `charge' rather than
the `potentials' before constructing the microcanonical partition function. If one
does this, then one finds that the corrected $\Psi^2$ tends to zero, rather than
infinity, as the CTCs are introduced. The situation is rather similar to that
encountered when trying to calculate the rate of pair production of electrically and
magnetically charged black holes. In that case, the introduction of a projection
operator is necessary to ensure that the pair production rates for both types of black 
holes are equal, as one would expect \cite{hawross}. In this case, it means that 
causality violating amplitudes are strongly suppressed, in accordance with the 
chronology protection conjecture.

\section{Acknowledgements}

The author would like to thank Stephen Hawking for valuable discussions. This work was
completed with the financial support of EPSRC.

\begingroup\raggedright\endgroup

\end{document}